# Three Other Models of Computer System Performance

Mark D. Hill, 28 December 2018
Computer Sciences Dept., University of Wisconsin-Madison

*This note argues for more use of simple models beyond
Amdahl's Law: Bottleneck Analysis, Little's Law, and a M/M/1 Queue.*

Computer systems from the Internet-of-Things devices to datacenters are complex, and optimizing them can enhance capability and save money. Existing systems can be studied with measurement, while prospective systems are most often studied by extrapolating from measurements of prior systems or via simulation software that mimics target system function and provides performance metrics. Developing simulators, however, is time-consuming and requires much infrastructure regarding a prospective system.

Analytic models—including simple ones like **Amdahl's Law**—represent a third underused evaluation method that can provide insight for both practice and research, albeit with less accuracy. Amdahl proposed his law [A67] to give an upper bound in the speedup of a N-processor computer system as $1/[(1-f) + f/N]$ where f is the fraction of work parallelizable. More generally, if a component X is used by a work fraction f and can be sped up by speedup $S_X$, Amdahl's Law gives an upper bound on overall speedup as $1/[(1-f) + f/S_X]$. While simple, Amdahl's Law has proven useful, including quickly showing what is not possible.

Our thesis is that analytic models can complement measurement and simulations to give quick insight, show what is not possible, provide a double-check, and suggest future directions. To this end, we present three simple models that we find useful like Amdahl's Law: **Bottleneck Analysis, Little's Law,** and **the M/M/1 Queue**. These can give initial answers to questions like:

1. What is the maximum throughput through several subsystems in series and parallel?
2. How many buffers are needed to track pending requests as a function of needed bandwidth and expected latency?
3. How can one both minimize latency and maximize throughput for unscheduled work?

These models are useful for insight regarding the basic computer system performance metrics of latency and throughput (bandwidth). Recall that *latency*—in units of time—is the time it takes to do a task (e.g., an instruction or network transaction). *Throughput* (called **bandwidth** in communication)—in units of tasks per time—is the rate at which tasks are completed. Throughput can be as simple as the reciprocal of latency but can be up to N times larger for N-way parallelism or N-stage pipelining.

## Bottleneck Analysis

Bottleneck Analysis [L84] can provide an upper bound on the throughput (bandwidth) of a system from the throughput of its subsystems. It provides a quick calculation for what a system may or can't do. If throughput is bottlenecked too low, no amount of buffering or other latency-reducing techniques can help.

Bottleneck Analysis gives the maximum throughput of a system from the throughputs of the system's subsystems with two rules that can be applied recursively:

1. The maximum throughput of K sub-systems in *parallel* is the *sum* of the subsystem throughputs.
2. The maximum throughput of K sub-systems in *series* is the *minimum* of the subsystem throughputs.

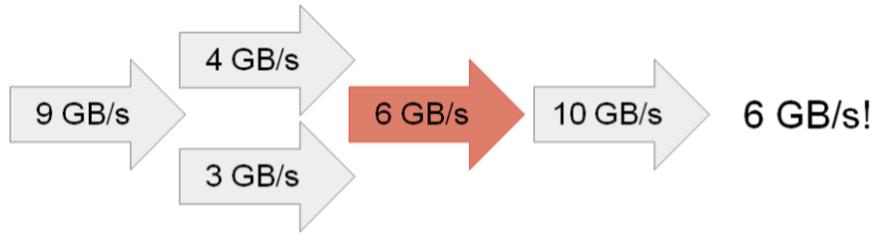

**Figure 1: Bottleneck Analysis (bottleneck is colored)**

Consider an example and literature pointers. Figure 1 gives an example where the arrows represent subsystems and bottleneck analysis finds the system's maximum throughput to be: 6 GB/s = MIN(9, (4+3), 6, 10). Roofline [W09] and Gables [H19] use bottleneck analysis to provide insight regarding multicore chips and Systems on a Chip (SoCs), respectively. Lazowska et al. [L84] (now online) more fully develops bottleneck analysis and other methods.

Bottleneck Analysis can also bring attention to system balance: decreasing all non-bottlenecked throughputs until they are (near) bottlenecks can reduce cost without changing throughput. Generally, systems should be bottlenecked by the most expensive subsystem(s) first.  However, Bottleneck Analysis ignores latency, which can be important as the next two models show.

**Little's Law**

Little's Law [L61] shows how total latency and throughput can relate. A colleague and I used to remark that Little's Law was our key industrial consulting tool. Consider a system in Figure 2 in steady state where tasks arrive, stay in the system, and depart. Let R be the rate (throughput, bandwidth) that tasks arrive (and, in steady state, they must depart at the same rate R), T be the average number of tasks in the system, and L be the average latency a task spends in the system. Little's Law says:

> Average number of tasks in system T = average latency L * average arrival rate R

For example, how many buffers must a cache have to record outstanding misses if it receives 2 memory references per cycle at 2.5 GHz, has miss ratio 6.25%, and average miss latency is 100 ns? Little's Law reveals the answer of 32 buffers.[1] However, as we will see later, more buffers will be needed for the common case when misses occur unscheduled and bursts make some miss latencies larger than 100 ns.

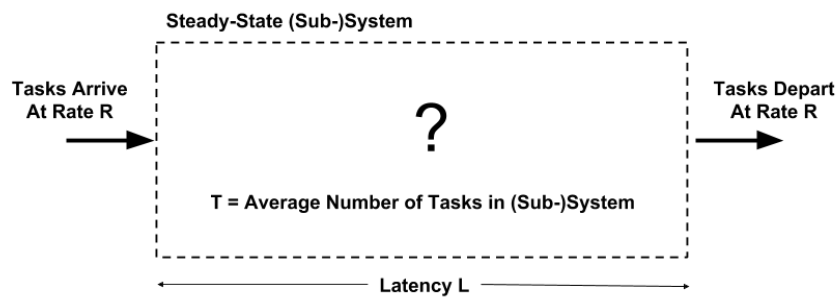

**Figure 2: (Sub-)System for Little's Law**

---

[1] L = 100 ns; R = 0.3125 Gmisses/s = (2 ref./cycle)*(2.5 GB/s)*(0.0625 misses/ref.); T = 32 ≥ 100*0.3125.

A beauty of Little's Law is that it is tri-directional: given any two of its parameters, one can calculate the third. This is important as some parameters may be easier to measure than others. Assume that your organization processes travel reimbursements and wants to know the average processing time (latency). If 200 requests arrive per day and there are 10,000 requests pending in the system, then Little's Law reveals an average of 50 days to process each request = 10,000 requests / (200 requests/day).

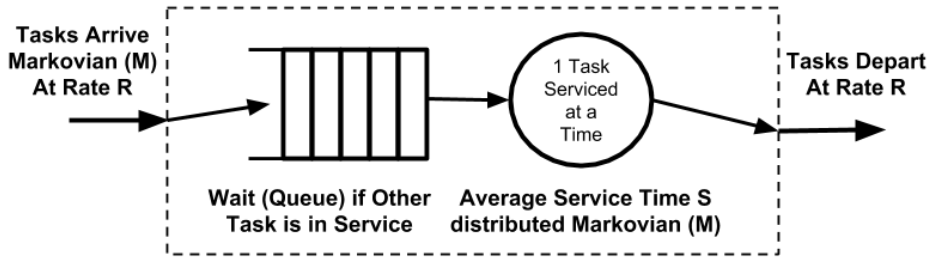

**Figure 3: M/M/1 Queue**

### A M/M/1 Queue

Little's Law provides a black-box result, but does not expose tradeoffs. This section's M/M/1 queue will show us a required trade-off among (a) allowing unscheduled task arrivals, (b) minimizing latency, and (c) maximizing throughput.

To be more precise, let's posit that "unscheduled task arrivals" means this property: **When a task arrives is independent (unaffected) of when previous tasks arrived.** For example, my user request doesn't know when you made yours. Queuing Theory [K75] has named this property "Markovian" (or Poisson arrivals or exponential inter-arrival times).

Figure 3 depicts a simple queue where tasks arrive, are optionally queued in the buffers to the left if the server is already busy, are then serviced one at a time in the circular server, and depart to the right. Let tasks arrive at rate R, optionally wait for average queuing time Q, and be serviced (without further queuing) at average time S.[2] Let L denoted the average total latency to handle a task, equal to Q + S.

Figure 3's queue becomes M/M/1 with three further assumptions: tasks arrive Markovian (1st M), are serviced Markovian (2nd M), and are serviced one at a time (final 1). The middle assumption—that tasks are serviced Markovian—seems strange but is an effective model of service time variability (whereas assuming all service times always equal S is usually optimistic).[3]

Figure 4 plots the M/M/1 queue latency-throughput tradeoff. The x-axis is the fraction of maximum throughput achieved x = R*S.[4] The y-axis is average total latency (L)—queuing plus service—normalized by dividing the no-waiting service time (S) for y = L/S. Thus, y = 1 means no waiting. Finally, queuing theory [K75] provides the plotted curve as L/S = 1/(1-R*S) or more simply y = 1/(1-x). Figure 4 reveals:

**High throughput ⇒ high latency:** As the fraction of maximum throughput approaches 1, latency explodes toward infinity, making full throughput infeasible. Intuitively, near-full throughput makes it harder to "recover" from a backlog of tasks after inevitable Markovian arrival bursts.

---

[2] We simplify from standard queueing theory notation that usually denotes R as $\lambda$ and 1/S as $\mu$.
[3] Also the arrival rate is not affected by the unbounded number of tasks in queue ("open system").
[4] This follows because actual throughput matches the arrival rate R, maximum throughput is 1/S as the reciprocal of one-at-a-time average service time S, and so the fraction of throughput is R/(1/S) = R*S.

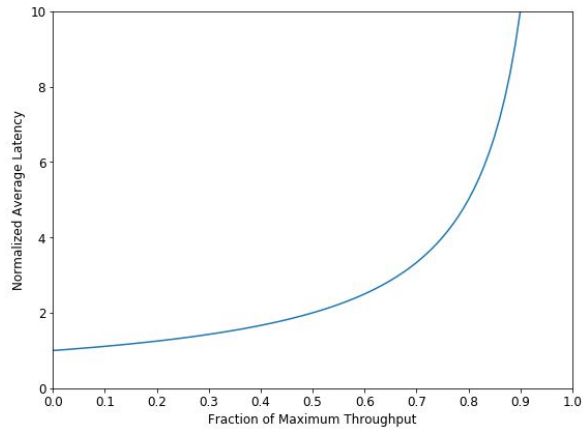

**Figure 4: M/M/1 Queue Latency vs. Throughput Tradeoff**

**Low latency ⇒ low throughput:** As throughput approaches 0, normalized latency approaches 1 where there is no queuing delay before service (i.e., arrivals rarely wait for previous tasks).

**Intermediate latency/throughput values reveal a stark trade off:** Throughput fractions 0.25, 0.50, 0.75, and 0.90 see normalized latencies of 1.33, 2.00, 4.00, and 10.00, respectively.

Consider an example applying a M/M/1 queue. Assume that a computer must handle network packets arriving unscheduled at a rate R = 10 packets/second with a goal of handling each packet with a total latency (including waiting) of L = 100 ms. How long can the computer take to process (service) each packet (not counting waiting), denoted S? A first answer might set S = 100 ms as that is the average time between packet arrivals.

For unscheduled arrivals, however, M/M/1 warns us that the above design will fail as maximal throughput causes exploding latency. Instead with M/M/1, we must service each packet in an average of S = 50 ms with 50 ms average delay for queuing and 50% of maximum throughput.[5] The M/M/1 design seems to over-designed by 2x (100ms → 50ms), but this is necessary.

For the previous cache miss buffer example, the 32-buffer answer is minimal for 100-ns average miss latency. Nevertheless, it would be prudent to increase buffering (e.g. another 50% to 48) to handle inevitable arrival bursts, unpredictable miss latency, or both.

On the other hand, to achieve low latency and highest bandwidth together, one must discard a M/M/1 queue and coerce task arrivals to not be Markovian and, in the best case, be scheduled (e.g., like at the dentist). Good online introductions to queuing theory include Lazowska et al. [L84] and Tay [T18].

Please keep these three "other" models in mind during the next presentation with numbers that seem too good to believe. Hill is supported by NSF CCF-1617824, NSF CNS-1815656, and John P. Morgridge Endowed Chair. We thank Dan Gibson, Swapnil Haria, Matt Sinclair, Guri Sohi and Daniel Sorin for improving this paper.

---

[5] Solve for S in L/S = 1/(1-R*S) when R = 10 packets/second and L = 100 ms.